\documentclass[twocolumn,prb,aps]{revtex4}
\usepackage{amsmath}
\usepackage{amssymb}
\usepackage{graphicx}
\usepackage{dcolumn}
\usepackage{graphicx}
\usepackage{dcolumn}
\usepackage{bm}

\begin{document}

\preprint{Phys.Rev.B}

\title{Aharonov Bohm effect in 2D topological insulator.}
\author{G.M.Gusev,$^1$ Z.D.Kvon,$^{2,3}$ O.A.Shegai,$^2$ N.N.Mikhailov,$^2$
 and S.A.Dvoretsky,$^{2}$}

\affiliation{$^1$Instituto de F\'{\i}sica da Universidade de S\~ao
Paulo, 135960-170, S\~ao Paulo, SP, Brazil}
\affiliation{$^2$Institute of Semiconductor Physics, Novosibirsk
630090, Russia} \affiliation{$^3$Novosibirsk State University,
Novosibirsk, 630090, Russia}

\date{\today}
\begin{abstract}
We present magnetotransport measurements in HgTe  quantum well with
inverted band structure, which expected to be a two-dimensional
topological insulator having the bulk gap with helical gapless
states at the edge. The negative magnetoresistance is observed in
the local and nonlocal resistance configuration followed by the
periodic oscillations damping with magnetic field. We attribute such
behaviour to Aharonov-Bohm effect due to magnetic flux through the
charge carrier puddles coupled to the helical edge states. The
characteristic size of these puddles is about 100 nm.


\end{abstract}

\maketitle
\section{Introduction}
The investigation of the quantum interference phenomena, such as
Aharonov-Bohm  (AB) oscillations and  weak localizations, provides
an important information about fundamental properties of various
electronic systems \cite{lee}. Recently new class of materials with
interesting properties have emerged, called topological insulators
(TI), which are insulating in the bulk and characterized by the
existence of robust gapless excitations at their surface
\cite{hasan, qi, moore, moore2}. Manifestation of AB oscillations in
topological insulators, can be utilized to probe the surface nature
in TI and its properties. The time-reversal-symmetric 2D topological
insulator is induced by a strong spin-orbit interaction \cite{kane,
bernevig, maciejko1, yang, chang1} and characterized by edge modes with opposite
spins propagating in opposite directions. The 2D TI have been
realized in HgTe quantum wells with inverted band structure
\cite{konig, buhmann}. Aharonov Bohm oscillations is intimately
related to the weak localization (WL) corrections. For example, WL
phenomena is discussed in terms of the collective action of the
magnetic flux through the random loops of the pair of time reversed
trajectories \cite{lee}. The constructive interference of these path
in return point is suppressed by AB flux, which lead to the
resistance increase and negative magnetoresistance \cite{altshuler}.
The spin orbit coupling strongly modifies the quantum  corrections
the conductivity, it leads to a destructive interference  between
clockwise and counter-clockwise trajectories and changes the WL to
weak anti-localization behavior (WAL).

In 2D topological system WL  is quite different from conventional 2D
metals and strongly affected by the Dirac spectrum of massive
fermions \cite{ostrovsky,tkachov} with the mass proportional to the
band gap. The electronic spectrum is shown in Figure 1. The band
structure has a simple  parabolic form near the bottom of the band
(marked by $E_{F}^{*}$ in Figure 1 and linear Dirac-like dispersion
relation at high Fermi energy $E_{F}$. When the Fermi energy lies in
the bulk gap near the charge neutrality point (CNP), spectrum
described by pair of the helical edge states (Figure 1). More over,
the edge state must have linear Dirac like dispersion. The
localization corrections are strongly governed by such remarkable
property of the spectrum. When the Fermi energy is small, the
effective spin-orbit coupling is weak, and one can expect the
conventional WL behaviour \cite{ostrovsky}. When the Fermi energy
becomes larger than the gap width, the energy dispersion is linear,
and theory describes crossover from WL to WAL behaviour
\cite{ostrovsky, tkachov, tkachov2, ostrovsky2}. The
magnetoresistance of 2D TI with a dominant edge state contribution
is described by two mechanisms: first scenario relies to the
frequent deviations of the edge electrons into the disordered AB
flux threaded 2D bulk \cite{maciejko}; and second scenario to the
localization of the helical edge sates due to the collective action
of the random magnetic flux through the loops, naturally formed by
rough edges of the sample \cite{deplace}. Both models predict
quasi-linear positive magnetoresistance MR. According to this
theoretical predictions positive MR was observed in 8 nm HgTe
quantum wells \cite{gusev2}.

\begin{figure}[ht!]
\includegraphics[width=9cm,clip=]{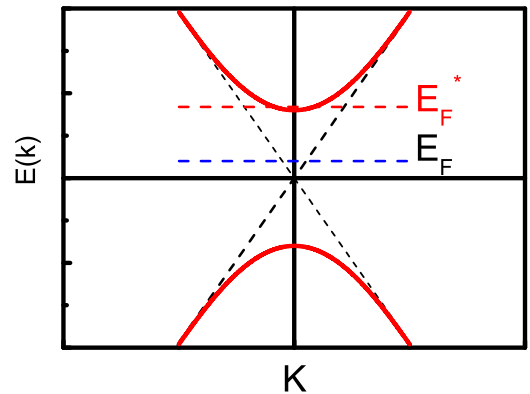}
\caption{\label{fig.1}(Color online)Schematic of the energy spectrum
of 2D TI near k=0. Solid lines -spectrum for the bulk electrons,
dashes- helical edge state spectrum. Positions of the Fermi energy
in the bulk gap and near the bottom of the conductive bands are
indicated. }
\end{figure}
Recently interaction of the helical states with multiple puddles of
charge carriers formed by fluctuations in the donor density has been
considered in 2D topological insulators \cite{vayrynen}. This leads
to significant inelastic backscattering due to the dwelling of
carriers in the puddles and to a large resistance, which depends
weakly on temperature. Moreover, carriers in the puddles describes
loops trajectories, which are sensitive to the magnetic field flux.
therefore, one could expect that the magnetoconductivity is mainly
contributed by the quantum interference of such electrons.

In the present paper we investigate transport properties of the HgTe
quantum wells with the width $d$ of 8-8.3 nm. Slightly above the the
charge neutrality point, when the system is expected to be
two-dimensional topological insulator, we observe the negative
magnetoresistance measured in the local and nonlocal configurations,
followed by periodic oscillations damping with magnetic field. When
the applied field is larger than 1 T, the MR becomes positive. From
the sign of the magnetoresistance we specified two contributions:
electrons from edge states and bulk electrons, which are localized
in the metallic puddles formed by  potential fluctuations.

\section{Experiment}

The $Cd_{0.65}Hg_{0.35}Te/HgTe/Cd_{0.65}Hg_{0.35}Te$ quantum wells
with (013) surface orientations and a width $d$ of 8-8.3 nm were
prepared by molecular beam epitaxy. A detailed description of the
sample structure has been given in \cite{kvon, olshanetsky}. Device
A is six-probe Hall bar, while device B is designed for
multiterminal measurements. The device A  was fabricated with a
lithographic length $6 \mu m$ and width $5 \mu m$ (Figure 2, top
panel). The device B consists of three $4 \mu m$ wide consecutive
segments of different length ($2, 8 , 32 \mu m$), and 7 voltage
probes. Device C (figure 5) is a structure with large gate area for
identifying nonlocal transport over macroscopic distances
\cite{gusev}. The lengths of the edge states are determined by the
perimeter of the sample part covered by metallic gate (mostly side
branches) rather than by the length of the bar itself. The ohmic
contacts to the two-dimensional gas were formed by in-burning of
indium. To prepare the gate, a dielectric layer containing 100 nm
$SiO_{2}$ and 200 nm $Si_{3}Ni_{4}$ was first grown on the structure
using the plasmochemical method. Then, the TiAu gate with sizes of
$18\times10 \mu m^{2}$ was deposited. The ungated HgTe well was
initially n-doped with density $n_{s}=1.8\times10^{11} cm^{-2}$.
Several devices with the same configuration have been studied. The
density variation with gate voltage was $1.09\times 10^{15}
m^{-2}V^{-1}$. The magnetotransport measurements in the structures
described were performed in the temperature range 1.4-25 K and in
magnetic fields up to 12 T using a standard four point circuit with
a 3-13 Hz ac current of 0.1-10 nA through the sample, which is
sufficiently low to avoid overheating effects.
\begin{figure}[ht!]
\includegraphics[width=8cm,clip=]{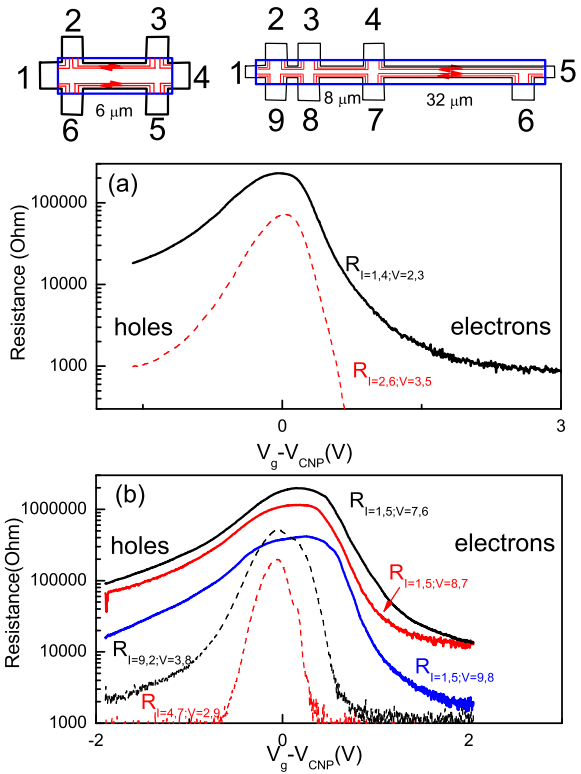}
\caption{\label{fig.2}(Color online) Color online) (a) a) The local
and nonlocal resistances $R$   as a function of the gate voltage at
zero magnetic field measured by various voltage probes for samples A
(a) and B (b), T=4.2 K. Top panel-shows schematic view of the
samples. The perimeter of the gate is shown by blue rectangle. }
\end{figure}
The density of the carriers in the HgTe quantum wells can be
electrically manipulated with local gate voltage $V_{g}$. The
typical dependence of the four-terminal resistance of two of the
representative samples A and B as a function of $V_{g}$ is shown in
Figure 2. The resistance $R_{14,23}=R_{I=1,4;V=2,3}$ of the sample A
and resistances for sample B, measured by various voltage probes in
a zero magnetic field reveal a sharp peaks, shown in figure 2 a and
b, when the gate voltage induces an additional charge density,
altering the quantum wells from an n-type conductor to a p-type
conductor via a 2D TI state. It has been shown \cite{konig, buhmann}
that the 4-probe resistance in an HgTe/CdTe micrometer-sized
ballistic Hall bar demonstrated a quantized plateaux
$R_{14,23}\simeq h/2e^{2}$. It is expected that the scattering
between the helical edge states in the topological insulator is
unaffected by the presence of a weak disorder \cite{kane, bernevig,
hasan}. Note, however, that the resistance of samples longer than $1
\mu m$  might be  much higher than $h/2e^2$ due to  the presence of
the electron spin flip backscattering on each boundary. Mechanism of
the back scattering is not clear and appealing task for
theoreticians and a matter of ongoing debate \cite{vayrynen,
maciejko2, strom}. The Hall effect reverses its sign and
$R_{xy}\approx0$ (not sown) when longitudinal resistance $R_{xx}$
approaches its maximum value, which can be identified as the charge
neutrality point (CNP). These behaviour is similar to those
described in graphene \cite{sarma}. An unambiguous way to prove the
presence of edge state transport mechanism in 2D TI with strong
backscattering on the boundary are the nonlocal electrical
measurements. The nonlocal response always exists because of the
presence of the two counter-propagating edge states, which flows
sideways and may reach any contacts in the device \cite{roth,
gusev}. Figures 2 a and b show the nonlocal resistances
corresponding to the different configurations. For example, nonlocal
resistance in figure 2a corresponds to the contact configuration,
when the current flows between contacts 2 and 6 and the voltage is
measured between contacts 3 and 5. One can see that the nonlocal
resistance
 near CNP has a peak of a comparable
amplitude, though less wide, and approximately in the same position
as the local resistance. Outside of the peak the nonlocal resistance
is negligibly small. The apparent residual nonlocal resistance in
figure 2b above $V_{g}- V_{CNP} > 1 V$ and below $V_{g}- V_{CNP} < -
1 V$ is is related to the noise in the full linear scale
measurements.

\begin{figure}[ht!]
\includegraphics[width=8cm,clip=]{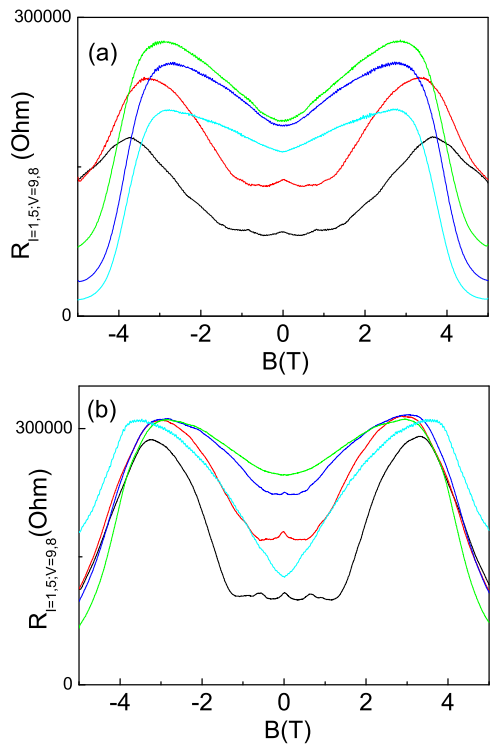}
\caption{\label{fig.3}(Color online) Color online) The local
resistance $R$ for two samples  with multiterminal configuration
(devices B) as a function of the magnetic field for the different
values of the gate voltages, when the Fermi energy passes through
CNP from the hole to electronic side of the resistance peak. (a)
$V_{g}-V_{CNP}$ (V): -0.15 (cyan), -0.02 (blue), 0.1 (green), 0.35
(red), 0.47 (black). (b)$V_{g}-V_{CNP}$ (V): -0.18 (cyan), 0.05
(green), 0.25 ( blue), 0.37 (red), 0.45 (black). }
\end{figure}

It is expected that a magnetic field perpendicular to the quantum
well breaks time reversal symmetry (TRS) and thereby enables elastic
scattering between counterpropagating chiral edge modes. However, a
number of the different theoretical models has previously been
proposed \cite{konig,tkachov,scharf,chen,maciejko2} with
 substantially different physical scenarios. In our previous study
 \cite{gusev2} we have observed a linear negative
magnetoconductance in HgTe-based quantum wells in the 2D TI regime
near CNP, when the edge state transport prevails. Our observation
agrees with the model \cite{maciejko2} which describes the effects
of WAL (see discussion section below for more details). The model
predicts almost linear positive magnetoresistance $\frac{\Delta
R}{e^{2}/h}=A|B|$, where parameter A strongly depends on the
disorder strength W in comparison with the energy gap $E_{g}$ : the
fluctuations $W>E_{g}$ result in a large B-slope corresponding to a
strong disorder regime and fluctuations $W<E_{g}$ lead  a small
B-slope. Figure 3 shows the evolution of the resistance $R_{xx}$
with magnetic field and density, when the chemical potential crosses
the bulk gap. The magnetoresistance (MR) demonstrates a striking
V-shape dependence in magnetic fields below 1T near CNP, which
confirms our previous observations. Note however, that the V-shaped
magnetoresistance is strongly transformed, when Fermi energy moves
away from CNP to the electronic side of the resistance peak. In this
region MR inverses the sign near zero magnetic field and shows
triangular-shaped peak accompanied by two satellite features or
damped oscillations. The width of the negative magnetoresistance
spike and the period of the oscillations are slightly varied from
sample to samples, as one can see in figures 3a and b. Both samples
are almost identical with approximately equal mobility. Note that
the Fermi level still lies in the bulk gap and transport is
dominated by edge states, because we see nonlocal effects. In the
hole side of the peak we do not find neither small negative
magnetoresistance near B=0, nor the oscillations on the side
branches of the the positive magnetoresistance. In magnetic fields
above 3 T the magnetoresistance falls off rapidly marking a
pronounced crossover to the quantum Hall effect regime in accordance
with previous observations \cite{gusev2}.

\begin{figure}[ht!]
\includegraphics[width=7cm,clip=]{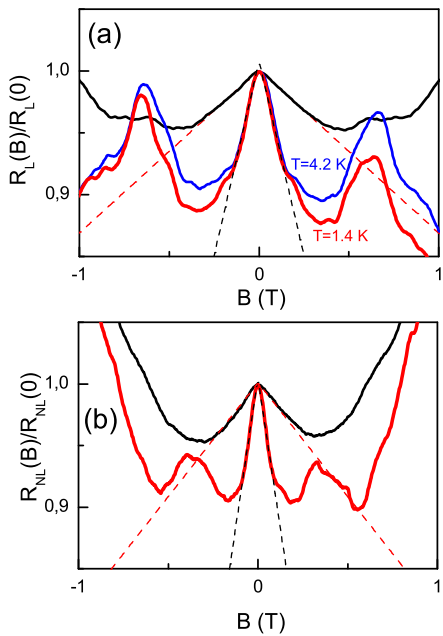}
\caption{\label{fig.4}(Color online) Color online) (a) The relative
local resistance $R_{L}(B)/R_{L}(0)$($R_{L}=R_{I=1,4;V=2,3}$) as a
function of the magnetic field for the two values of the gate
voltage $V_{g}-V_{CNP}$(V): 0.35 (black), 0.53 (blue, red), and two
temperatures (device B). (b) The relative nonlocal resistance
$R_{NL}(B)/R_{NL}(0)$ ($R_{NL}=R_{I=2,6;V=3,5}$) as a function of
the magnetic field for the two values of the gate voltage
$V_{g}-V_{CNP}$ : 0.35 (black), 0.5 (red) (device A). Dashed lines
are B-linear approximations. }
\end{figure}

The figure 4 shows the low-fled part of the relative
magnetoresistance  for the two values of the gate voltages and two
temperatures. One can see that for this particular sample the MR
varies linearly with magnetic field. As voltage $V_{g}$ increases,
the B-slope of the MR decreases and additional oscillation
emerges. The MR profile does not show any significant temperature
dependence. Figure 4 b displays traces of the resistance in
nonlocal configuration for device B ($R_{NL}=R_{I=2,6;V=3,5}$).
One can see similar triangular-shaped MR peak, as in the local
geometry, though less wide, with two satellite peaks. Coexistence
of the low-field negative MR peak in nonlocal configuration
exclude the possibility that this effect has a bulk origin. It is
worth noting , however, that around B=0 the magnetoresistance is
parabolic rather than linear in nature.
\begin{figure}[ht!]
\includegraphics[width=8cm,clip=]{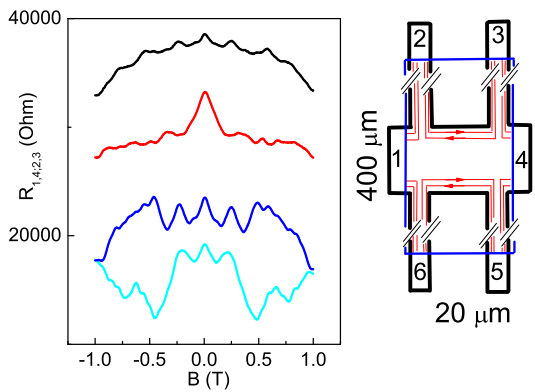}
\caption{\label{fig.5}(Color online) Color online) The local
resistance $R$ for device C as a function of the magnetic field for
the different values of the gate voltages. (a) $V_{g}-V_{CNP}$ (V)
(from top to bottom): 0.1, 0.3, 0.5, 0.7. }
\end{figure}

Finally we present the results for a device C with a large gate. We
expect that the length of the edge states in this structure is
determined by the perimeter of the sample covered by the metallic
gate. Figure 5 shows schematic view of the sample. One can see that
the current flows mostly along the edges of the side branches. The
resistance reveals saw-tooth oscillations, shown in Fig.5, when the
Fermi energy passes the electronic side of the resistance peak away
from CNP. Note, however, that the evolution of the oscillations with
gate occurs rapidly and not gradually, as in devices A and B.

\section{Discussion}

We now proceed to an analysis of the data described. We will focus
on the two models that can explain the magnetoresistance of one
dimensional edge electrons. Both models considered alternative paths
for the edge states due to bulk disorder or rough edges of the
realistic sample. Figure 6 illustrates the helical edge states in a
disordered 2D TI in an uniform magnetic field. The first scenario
describes disordered spinless one dimensional quantum wire
\cite{maciejko2}. For strong enough disorder electron paths
frequently deviate into the bulk region enclosing AB flux before
returning back to the edge, labeled in figure 6 by number 2. The
conventional WAL approach can be used. The average with respect to
the different size of the loops leads to the positive linear
magnetoresistance. Second scenario relies to the localization of the
helical edge sates due to the collective action of the random
magnetic flux through the loops, naturally formed by rough edges of
the sample  (labeled  by number 1 in the Figure 6) \cite{deplace}.
In accordance with this scenario immediately after magnetic field is
switched on, edge states become localized. The magnetic field
penetrates through the random helical edge states loops (Figure 6 ),
which acts as magnetic flux impurity and introduces the
backscattering between edge states on each boundary. Similar to the
first scenario the average with respect to the different magnetic
fluxes should be performed. The model \cite{deplace} predicts
positive magnetoresistance and $B^{2}$ dependence of the inverse
localization length $\gamma$ in small magnetic field, which becomes
more linear with increasing magnetic field. Our experimentally
observed linear positive magnetoresistance near CNP are in good
qualitative agreement with both models. Detailed comparison with
model \cite{maciejko2} has been performed in our previous
publication \cite{gusev}. Discrimination among  two scenarios
requires further experimental work. It is worth noting that  both
models  consider ballistic transport at zero magnetic field, while
our samples demonstrate diffusive transport. Despite the fact that
while the both models give a satisfactory description of the linear
positive magnetoresistance near CNP, the explanation of negative MR
and  AB-like oscillations in Fig.3-4 away from CNP requires a
further elaboration of the models.
\begin{figure}[ht!]
\includegraphics[width=9cm,clip=]{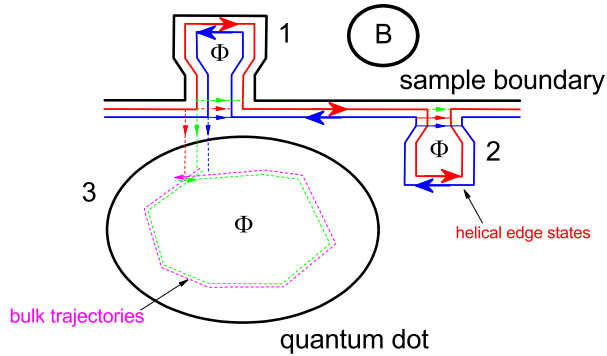}
\caption{\label{fig.1}(Color online)  Schematics of helical edge
state propagation along a disordered edge in 2DTI in a perpendicular
magnetic field. Electrons tunnel in and out quantum dot (puddle)
created by inhomogeneous charge distribution. Electron helical
trajectories form loops resulted from the edge roughness (1), near
bulk disorder (2) and metallic puddles (3). Bulk electron
trajectories form loops inside of the puddle.}
\end{figure}

Absence of the resistance quantization in samples with dimensions
above a few microns \cite{konig, buhmann, gusev} is another unsolved
problem. One of the possible explanation is the fluctuations of the
local insulating gap width induced by smooth inhomogeneities, which
can be represented as metallic puddles or dots. The well-localized
metallic regions along the edge have been found using scanning gate
microscopy \cite{konig2} and in microwaves experiments \cite{kvon2}.
According to the model \cite{vayrynen} charge carrier puddles
coupled to a coherent conductor results in incoherent inelastic
processes and modifies the ballistic transport. Therefore metallic
puddles can lead to spin dephasing, since an electron entering the
puddle is thermalized by dissipation and later on fed back into the
system. Therefore ballistic coherent transport is expected only in
the region between the puddles, and total 4-terminal resistance
exceeds the quantized value. Self-averaging resistance of the sample
with edge state dominated contribution to transport is given by
\cite{vayrynen}:
\begin{equation}\label{1}
R\sim
\frac{h}{e^{2}}\frac{1}{g^{2}}n_{p}\lambda\left(\frac{T}{\delta}\right)^{3}L
\end{equation}
where $n_{p}$ is the density of the puddles, $\lambda=\hbar
v/E_{g}\approx 18 nm$ is the electron penetration depth into the
puddles ($v \approx 5.5\times10^{7}cm/s$ is the electron velocity,
$E_{g}\simeq 20meV$ is the forbidden gap), g is the dimensionless
conductance within the dot (puddle), $\delta$ is the mean level
spacing within the dot, L is the distance between probes (length of
the edge states). It has been found  that the equation 1 gives a
satisfactory explanation  of the high resistance value, obtained in
experiments \cite{gusev3}. Combining all parameters we  calculate
$\rho_{0}=R/(\frac{h}{e^{2}}L)=8\times10^{3}\left(\frac{T}{\delta}\right)^{3}(e^{2}/h)/cm$,
which is comparable with experimental value
$\rho_{0}=15\times10^{3}
(e^{2}/h)/cm$. The density of the puddles
increases, and the size of the puddles is growing, when the Fermi
level lies nearer to the conductive band. The Figure 6 illustrates
the helical edge states in the presence of the all types of the
disorder: edge roughness (1), bulk disorder (2) and metallic puddles
due to inhomogeneous charge distributions. We ague that, while the
edge roughness as well as the bulk disorder are important and play a
dominant role in the region near CNP in magnetic field, away from
CNP the role of the metallic puddles in MR becomes more pronounced
and essential. WL effects in ballistic cavities indeed has been
studied both theoretically \cite{baranger} and experimentally
\cite{chang2}. The WL peak and periodic oscillations are observed
which were attributed to the Aharonov-Bohm effect through a periodic
orbit within the cavities. Therefore, it is naturally to explain the
negative magnetoresistance and AB-like oscillations in figures 3,4
by existence of metallic ballistic puddles near the edge. It is
worth noting that the Fermi energy of electrons in puddles lies in
the parabolic part of the energy spectrum ( figure 1) because of the
low density. Therefore, one would expect that the spin orbit
coupling is weak, and magnetoresistance is negative in agreement
with our observations. It is easy to estimate the characteristic
sizes of these loops using the period of the oscillations in
fig.3-4, which is $\delta B = 0.3-0.4$ T. Then the characteristic
area of the loops is $S = \delta B/\Phi_{0} (\Phi_{0}= h/e) =
10^{-10} cm^{2}$ and respectively size is about 100 nm. Indeed this
value agrees with estimations of the puddle size and density
\cite{gusev3} and scanning gate microscopy \cite{konig2}. Note that
for explanation of the negative sign of the MR the role of the bulk
electrons in the puddles is emphasized, however the bulk state and
edge state may co-exist (figure 6). The interplay between the
topological insulators helical states and bulk electrons requires
further theoretical study. When the Fermi level moves to valence
band, it is expected that the puddles should be occupied by the
holes. Note however, that both dephasing  and spin relaxation times
for holes are found to be much smaller than for electrons in similar
conditions and WAL effect should be considerably smaller in
agreement with our observations.

When the Fermi level lies in the conductive band, and transport
becomes dominant by massive fermions with Dirac spectrum (figure 1),
the weak antilocalization behaviour has been observed
\cite{olshanetsky2}.

In conclusion, we observed interplay between positive and negative
magnetoresistance, when Fermi level shifts with respect to the
charge neutrality point, but still lies inside of the gap, and
transport occurs via edge states. We consider three contributions to
the magnetoresistance: edge state penetration to the bulk, edge
state scattering by magnetic flux formed by rough edges, and
 WL of the bulk electrons in the puddles  formed by inhomogeneous charge
 distributions. The negative magnetoresistance and AB-like
 oscillations are attributed to weak localization of the bulk electrons
in the metallic puddles formed by fluctuation of the local
insulating gap.

We thank O.K.Raichev for helpful discussions. A financial support of
this work by FAPESP, CNPq (Brazilian agencies), RFBI and RAS
programs "Fundamental researches in nanotechnology and
nanomaterials" and "Condensed matter quantum physics" is
acknowledged.


\begin{references}

\bibitem{lee}
P. A. Lee, T. V. Ramakrishnan, Rev.Mod.Phys.{\bf 57}, 287 (1985).
\bibitem{hasan}
M. Z. Hasan, C. L. Kane, Rev.Mod.Phys. {\bf 82}, 2045 (2010); X-L.
Qi, S-C. Zhang,Rev.Mod.Phys. {\bf 83}, 1057 (2011)
\bibitem{qi}
 X-L. Qi, S-C. Zhang, Phys.Today, Phys. Today {\bf 63(1)}, 33 (2010).
\bibitem{moore}
 J. E. Moore and L. Balents, Phys. Rev. B {\bf 75} 121306 (2007)
\bibitem{moore2}
J. E. Moore, Nature (London) {\bf 464}, 194(2010).
\bibitem{kane}
C. L. Kane and E. J. Mele, Phys. Rev. Lett. {\bf 95}, 146802 (2005).
\bibitem{bernevig}
B. A. Bernevig, T. L. Hughes, and S. C. Zhang, Science {\bf 314},
1757 (2006).
\bibitem{maciejko1}
J.Maciejko,T. L. Hughes,and S-C Zhang, Annu. Rev. Condens. Matter
Phys. {\bf 2}, 31 (2011).
\bibitem{yang}
Wen Yang and Kai Chang,
Shou-Cheng Zhang, Phys.Rev.Lett. {\bf 100}, 056602 (2008).
\bibitem{chang1}
Kai Chang and Wen-Kai Lou, Phys. Rev.Lett. {\bf 106}, 206802 (2011).
\bibitem{konig}
M. K\"{o}nig \textit{et al}, Science {\bf 318}, 766 (2007).
\bibitem{buhmann}
H.Buhmann, Journal. Appl.Phys.,{\bf 109}, 102409 (2011).
\bibitem{altshuler}
B. L. Altshuler, D. Khmel'nitzkii, A. I. Larkin and P. A. Lee,
Phys. Rev. B {\bf 22}, 5142 (1980).
\bibitem{ostrovsky}
P. M. Ostrovsky, I. V. Gornyi, and A. D. Mirlin, Phys. Rev. Lett.
{\bf 105}, 036803 (2010).
\bibitem{tkachov}
G. Tkachov and E. M. Hankiewicz, Phys. Rev. B {\bf 84}, 035444
(2011).
\bibitem{tkachov2}
G. Tkachov, Phys. Rev. B, {\bf 88}, 205404 (2013).
\bibitem{ostrovsky2}
P. M. Ostrovsky, I. V. Gornyi, and A. D. Mirlin, Phys. Rev. B {\bf
86}, 125323 (2012).
\bibitem{maciejko}
J. Maciejko, X-L. Qi, and S-C. Zhang, Phys. Rev. B {\bf 82}, 155310
(2010).
\bibitem{deplace}
P. Delplace, J. Li, and M. B\"{u}ttiker, Phys. Rev. Lett. {\bf 109},
246803 (2012).
\bibitem{gusev2}
G. M. Gusev, E.B.Olshanetsky, Z. D. Kvon, N. N. Mikhailov and S. A.
Dvoretsky, Phys. Rev. B {\bf 87}, 081311(R), (2013).


\bibitem{vayrynen}
 J.I.Vayrynen, M.Goldstein, L.I.Glazman, Phys.Rev.Lett.
110, 216402 (2013).
\bibitem{kvon}
Z. D. Kvon, E. B. Olshanetsky, D. A. Kozlov, et al., Pis'ma Zh.
Eksp. Teor. Fiz. {\bf 87}, 588 (2008) [JETP Lett. {\bf 87}, 502
(2008)].
\bibitem{olshanetsky}
E. B. Olshanetsky, Z. D. Kvon, N. N. Mikhailov, E.G. Novik, I. O.
Parm, and S. A. Dvoretsky, Solid State Commun. {\bf 152}, 265
(2012).
\bibitem{gusev}
G. M. Gusev, Z. D. Kvon, O. A. Shegai, N. N. Mikhailov, S. A.
Dvoretsky, and J. C. Portal, Phys. Rev. B {\bf 84}, 121302(R),
(2011).

\bibitem{maciejko2}
J. Maciejko, C. X. Liu, Y. Oreg, X. L. Qi, C. Wu, and S. C. Zhang,
Phys. Rev. Lett. {\bf 102}, 256803 (2009).

\bibitem{strom}
A.Str\"{o}m, H.Johannesson, G.I.Japaridze, Phys. Rev. Lett. {\bf
104}, 256804 (2010).

\bibitem{sarma}
S. Das Sarma, Shaffique Adam, E. H. Hwang, Enrico Rossi, Rev. Mod.
Phys., {\bf 83}, 407 (2011).
\bibitem{roth}
A. Roth, C. Br$\ddot{u}$ne, H. Buhmann, L.W. Molenkamp, J. Maciejko,
X.-L. Qi, and S.-C. Zhang, Science {\bf 325}, 294 (2009).

\bibitem{scharf}
B. Scharf, A. Matos-Abiague, and J. Fabian, Phys. Rev. B {\bf 86},
075418 (2012).
\bibitem{chen}
J.C. Chen, J. Wang, and Q.F. Sun, Phys. Rev. B {\bf 85} 125401
(2012).
\bibitem{konig2}
M. K$\ddot{o}$onig, M. Baenninger, A. G. F. Garcia et al., Phys.
Rev. X {\bf 3}, 021003 (2013).
\bibitem{kvon2}
Z.D.Kvon et al, Pisma v ZhETF,{\bf 99}, 333 (2014).
\bibitem{gusev3}
G.M.Gusev, Z.D.Kvon,E.B.Olshanetsky, A.D.Levin, Y. Krupko, J. C.
Portal, N.N.Mikhailov,  and S.A.Dvoretsky, Phys. Rev. B {\bf 89},
125305 (2014).

\bibitem{olshanetsky2}
E. B. Olshanetsky, Z. D. Kvon, G.M.Gusev, et al, JETP
Lett,{\bf91}, 347 (2010).
\bibitem{baranger}
H. U. Baranger, R. A. Jalabert, and A. D. Stone, Phys. Rev. Lett.
{\bf 70}, 3876 (1993).

\bibitem{chang2}
A. M. Chang, H. U. Baranger, L. N. Pfeiffer, and K.W. West, Phys.
Rev. Lett. {\bf 73}, 2111 (1994).

\end{references}
\end{document}